\documentclass[a4paper]{IEEEtran}
\usepackage{graphicx}
\usepackage{amssymb,amsmath,amsthm,amsfonts}
\usepackage{booktabs}
\usepackage{multirow}
\usepackage{cite}
\usepackage{xcolor}
\usepackage{microtype}

\theoremstyle{remark}

\newcommand{\Red}[1]{{\color{red}#1}}
\renewcommand{\Red}[1]{{}}

\DeclareMathOperator{\diag}{diag}
\DeclareMathOperator{\trace}{Tr}

\title{Eigenvalues of  Symmetric Non-normalized\\ Discrete Trigonometric Transforms}
\author{Ali~Bagheri~Bardi,~
Milo\v{s}~Dakovi\'{c}~\IEEEmembership{Member,~IEEE},
Taher ~Yazdanpanah,
Ljubi\v{s}a~Stankovi\'{c},~\IEEEmembership{Fellow,~IEEE}
\thanks{A. Bagheri-Bardi (bagheri@pgu.ac.ir) is with Persian Gulf University, Bushehr, Iran, on sabbatical leave at the University of Montenegro}
\thanks{T. Yazdanpanah (yazdanpanah@pgu.ac.ir) is with Persian Gulf University, Bushehr, Iran}
\thanks{M. Dakovi\'{c} (milos@ucg.ac.me) and L. Stankovi\'{c} (ljubisa@ucg.ac.me) are with University of Montenegro, Faculty of Electrical Engineering, Podgorica, Montenegro. This paper is submitted to an IEEE journal.
}
}

\begin{document}

\maketitle

\begin{abstract}
A comprehensive approach to the spectrum characterization (derivation of eigenvalues and the corresponding multiplicities) for non-normalized, symmetric discrete trigonometric transforms (DTT) is presented in the paper. Eight types of the DTT are analyzed. New explicit analytic expressions for the eigenvalues, together with their multiplicities, for the cases of three DTT (DCT$_{(1)}$, DCT$_{(5)}$, and DST$_{(8)}$), are the main contribution of this paper. Moreover, the presented theory is supplemented by new, original derivations for the closed-form expressions of the square and the trace of analyzed DTT matrices.
\end{abstract}

\begin{IEEEkeywords}
Discrete trigonometric transforms, Discrete cosine transforms, Discrete sine transforms, eigenvalues
\end{IEEEkeywords}

\section{Introduction}
\IEEEPARstart{D}{iscrete} Trigonometric Transforms (DTT) are irreplaceable tools in signal and image processing applications. There exist 16 types of the DTT \cite{Mitra,LjubisaDSP,Strang1999,Puschel2003,Puschel2002,Britanak2010,Ahmed1974,Ochoa2019,Garcia2018}
divided into two classes: Discrete Cosine Transforms (DCT) and Discrete Sine Transforms (DST). In each class, eight types of these transforms are defined. In addition, there are non-normalized and normalized variants of the DTT. All of these transforms are linear, and therefore, for a given signal of length $n$, they can be suitably represented using $n\times n$ transformation matrices.

Herein, we will focus on the symmetric non-normalized DTT, that is, on the DST of type 1, 4, 5, and 8 and the DCT of type 1, 4, 5, and 8. The elements of the transformation matrix for each analyzed DTT are given in Table \ref{DTT-def}, where $k=0,1,\ldots,n-1$ is a row index and $l=0,1,\ldots,n-1$ is a column index. The DTT of type $m$ are denoted as  DCT$_{(m)}$ and DST$_{(m)}$, and the corresponding transformation matrices are denoted as $\mathbf{C}_{(m)}$ and $\mathbf{S}_{(m)}$.

Eigenvalues of the DTT are studied in \cite{Tseng2002,Pei2001,Pei2004,Candan2011,Wei2014,Cariolaro2002,Dickinson82}. However, analytic results are provided only for 5 types of symmetric non-normalized DTT, covering the cases when the square of the transformation matrix $\mathbf{A}$ is proportional to the identity matrix, $\mathbf{A}^2=\lambda \mathbf{I}$.

Eigendecompositions of the DST$_{(4)}$ and the DCT$_{(4)}$ are analyzed in \cite{Tseng2002}. Therein, the authors use the Generalized Discrete Fourier Transform (GDFT), and the theory of commuting matrices in order to obtain approximate eigendecompositions of DST$_{(4)}$ and DCT$_{(4)}$.
In \cite{Pei2001} it has been shown that the DCT$_{(1)}$ and DST$_{(1)}$ eigenvectors can be
attained from the DFT eigenvectors. 
The offset Discrete Fourier Transform (DFT) is used in \cite{Pei2004}, where it has been shown that an even-order DCT$_{(4)}$, DST$_{(4)}$, and DST$_{(8)}$ can be viewed as a special case of an even-order offset DFT. This approach has led to the eigenvalues (and their corresponding multiplicities) for these three types of DTT.
The approach based on commuting matrices  is used in \cite{Dickinson82,Wei2014} to determine the eigenvectors of some DTT. Non-symmetric DTT are analyzed in \cite{Cariolaro2002},  providing a conjecture that all eigenvalues are distinct for non-symmetric DTT of arbitrary order.

Our aim is to find the eigenvalues, with their corresponding multiplicities, in an analytic way, for each considered DTT.
Applying some well-known trigonometric identities, we directly obtain the square and the trace of all eight types of  DTT matrices. We observe that the square of the transformation matrix for three types of non-normalized DTT, (the DCT$_{(1)}$, DCT$_{(5)}$, and DST$_{(8)}$) is not a multiple of the identity matrix. Using the formula for the trace, we compute the multiplicity of the eigenvalues in all considered cases.

Herein, we develop a unified analytic approach to DTT eigenvalues (and corresponding multiplicities), containing novel results for the DCT$_{(1)}$, DCT$_{(5)}$, and DST$_{(8)}$. Mathematically relevant expressions for the square and the trace of the analyzed DTT matrices arise as intermediate results, that are used for the calculation of the eigenvalues and their corresponding multiplicities.

\begin{table}[tb]
\centering
\caption{Definitions for symmetric, non-normalized DTT}
\label{DTT-def}

\begin{tabular}{ccccc}
\toprule
Type & $\left(s_{kl}\right)_{k,l=0}^{n-1}$ & & Type & $\left(c_{kl}\right)_{k,l=0}^{n-1}$ \\
\cmidrule{1-2}\cmidrule{4-5}
DST$_{(1)}$ & $\sin\frac{(k+1)(l+1)\pi}{n+1}$ & & DCT$_{(1)}$ & $\cos\frac{kl\pi}{n-1}$ \\[2ex]
DST$_{(4)}$& $\sin\frac{(2k+1)(2l+1)\pi}{4n}$ & & DCT$_{(4)}$& $\cos\frac{(2k+1)(2l+1)\pi}{4n}$ \\[2ex]
DST$_{(5)}$ & $\sin\frac{2(k+1)(l+1)\pi}{2n+1}$ & & DCT$_{(5)}$ & $\cos\frac{2kl\pi}{2n-1}$ \\[2ex]
DST$_{(8)}$ & $\sin\frac{(2k+1)(2l+1)\pi}{4n-2}$ & & DCT$_{(8)}$ & $\cos\frac{(2k+1)(2l+1)\pi}{4n+2}$ \\
\bottomrule
\end{tabular}
\end{table}

The main results are summarized in Section \ref{sec-Results}. Squares of DTT matrices are evaluated in Section \ref{sec-squares} while their traces are derived in Section \ref{sec-traces}. In Section \ref{sec-Proofs}, the analytic proofs for the results presented in Section \ref{sec-Results} are presented.

\section{Results}
\label{sec-Results}
The eigenvalues along with their corresponding multiplicities, for the considered types of DTT are presented in Tables \ref{tab-ev} and \ref{tab-DCT1}. These expressions are the main result of this paper. 

By carefully observing these forms, we can see that five DTT types have only two distinct eigenvalues and that for an odd $n$, in each case, the multiplicity of the positive eigenvalue is greater by one than the multiplicity of the corresponding negative eigenvalue.

\begin{table}[tb]
\centering
\caption{Eigenvalues and corresponding multiplicities}
\label{tab-ev}
\begin{tabular}{clcc}
\toprule
\multirow{2}{*}[-1ex]{DTT type}  & \multirow{2}{*}[-1ex]{Eigenvalue} & \multicolumn{2}{c}{Multiplicity} \\
\cmidrule{3-4}
& & even $n$ & odd $n$ \\
\midrule
\multirow{2}{*}[-2.5ex]{DCT$_{(4)}$, DST$_{(4)}$} 
&  $\lambda_1=-\sqrt{\dfrac{n}{2}}$ & $\dfrac{n}{2}$  & $\dfrac{n-1}{2}$ \\
\cmidrule{2-4}
& $\lambda_2=\sqrt{\dfrac{n}{2}}$ & $\dfrac{n}{2}$ & $\dfrac{n+1}{2}$ \\
\midrule
\multirow{2}{*}[-2.5ex]{DCT$_{(8)}$, DST$_{(5)}$} & $\lambda_1=-\sqrt{\dfrac{2n+1}{4}}$ & $\dfrac{n}{2}$  & $\dfrac{n-1}{2}$ \\
\cmidrule{2-4}
& $\lambda_2=\sqrt{\dfrac{2n+1}{4}}$ & $\dfrac{n}{2}$  &  $\dfrac{n+1}{2}$ \\
\midrule
\multirow{2}{*}[-2.5ex]{DST$_{(1)}$} & $\lambda_1=-\sqrt{\dfrac{n+1}{2}}$ & $\dfrac{n}{2}$  &  $\dfrac{n-1}{2}$ \\
\cmidrule{2-4}
& $\lambda_2=\sqrt{\dfrac{n+1}{2}}$ & $\dfrac{n}{2}$  &  $\dfrac{n+1}{2}$ \\
\midrule
& $\lambda_1=\dfrac{1}{4}-\sqrt{n-\dfrac{7}{16}}$ & $1$  & $1$ \\
\cmidrule{2-4}
\multirow{2}{*}[-2ex]{DCT$_{(5)}$} & $\lambda_2=-\sqrt{\dfrac{2n-1}{4}}$ & $\dfrac{n}{2}-1$  &  $\dfrac{n-3}{2}$ \\
\cmidrule{2-4}
& $\lambda_3=\sqrt{\dfrac{2n-1}{4}}$ & $\dfrac{n}{2}-1$  &  $\dfrac{n-1}{2}$ \\
\cmidrule{2-4}
& $\lambda_4=\dfrac{1}{4}+\sqrt{n-\dfrac{7}{16}}$ & $1$  & $1$ \\
\midrule
& $\lambda_1=-\dfrac{(-1)^{n}}{4}-\sqrt{n-\dfrac{7}{16}}$ & $1$  & $1$ \\
\cmidrule{2-4}
\multirow{2}{*}[-2ex]{DST$_{(8)}$} 
& $\lambda_2=-\sqrt{\dfrac{2n-1}{4}}$ & $\dfrac{n}{2}-1$  & $\dfrac{n-3}{2}$ \\
\cmidrule{2-4}
& $\lambda_3=\sqrt{\dfrac{2n-1}{4}}$ & $\dfrac{n}{2}-1$  &  $\dfrac{n-1}{2}$ \\
\cmidrule{2-4}
& $\lambda_4=-\dfrac{(-1)^{n}}{4}+\sqrt{n-\dfrac{7}{16}}$ & $1$  & $1$ \\
\bottomrule
\end{tabular}
\end{table}

For the DCT of type 5 and the DST of type 8 there are four eigenvalues. Two of them have multiplicity one, and the multiplicities of other two eigenvalues are equal in the case of even $n$, or differ by one in the case of odd $n$, where again the positive eigenvalue has greater multiplicity.

The most complicated  case is the DCT of type 1, where there are six distinct eigenvalues. Four of them have multiplicity one and the remaining two eigenvalues have equal multiplicities (for even $n$), whereas the positive eigenvalue multiplicity is greater by one than the  negative eigenvalue multiplicity (for odd $n$ case). 

In the following section, we provide detailed discussion, derivations and proofs for the presented results.

\begin{table}[tb]
\centering
\caption{DCT type 1 eigenvalues}
\label{tab-DCT1}
\begin{tabular}{lclc}
\toprule
\multicolumn{2}{c}{odd $n$} & \multicolumn{2}{c}{even $n$} \\
\cmidrule(lr){1-2} \cmidrule(lr){3-4}
\multicolumn{1}{c}{Eigenvalue} & Mult. & \multicolumn{1}{c}{Eigenvalue} & Mult. \\
\midrule
$\lambda_1=-\sqrt{n-1}$ & $1$ & $\lambda_1=-\dfrac{\sqrt{2}}{4} -\sqrt{n-\dfrac{7}{8}}$ & 1 \\
\midrule
$\lambda_2=\dfrac{1}{2} -\sqrt{n-\dfrac{3}{4}}$ & $1$ & $\lambda_2=\dfrac{\sqrt{2}}{4} -\sqrt{n-\dfrac{7}{8}}$ & 1  \\
\midrule
$\lambda_3=-\sqrt{\dfrac{n-1}{2}}$ & $\dfrac{n-5}{2}$ & $\lambda_3=-\sqrt{\dfrac{n-1}{2}}$ & $\dfrac{n}{2}-2$ \\
\midrule
$\lambda_4=\sqrt{\dfrac{n-1}{2}}$ & $\dfrac{n-3}{2}$ & $\lambda_4=\sqrt{\dfrac{n-1}{2}}$ & $\dfrac{n}{2}-2$ \\
\midrule
$\lambda_5=\sqrt{n-1}$ & $1$ & $\lambda_5=-\dfrac{\sqrt{2}}{4} +\sqrt{n-\dfrac{7}{8}}$ & 1 \\
\midrule
$\lambda_6=\dfrac{1}{2} +\sqrt{n-\dfrac{3}{4}}$ & $1$ & $\lambda_6=\dfrac{\sqrt{2}}{4} +\sqrt{n-\dfrac{7}{8}}$ & 1  \\
\bottomrule
\end{tabular}
\end{table}

\section{Square of transformation matrix}
\label{sec-squares}
In order to compute entries $t_{kl}$ of the square of a DTT matrix, we need to apply some well-known trigonometric identities \cite[p.37]{TrigBook}.

Lagrange's trigonometric identity states that for $\theta\ne 2k\pi$, where $k$ is an integer, holds
$$
\sum_{m=0}^{n} \cos m\theta=\frac{1}{2} + \frac{\sin(n+\frac{1}{2})\theta}{2\sin \frac{\theta}{2}}.
$$
Using this identity for $\theta=(a\pi)/n$,  where $a$ is an integer that is not divisible with $2n$, and for $\theta=(2b\pi)/(2n+1)$, where $b$ is an integer not divisible with $2n+1$ , we get the following two identities
\begin{gather}
\sum_{m=0}^n\cos \frac{m a\pi}{n}  =\frac{1+(-1)^a}{2} =
\begin{cases}
0 & \text{for odd } a \\
1 & \text{for even } a
\end{cases}
 \label{Lag1} \\ 
 \sum_{m=0}^n\cos \frac{2mb \pi}{2n+1}  =\frac{1}{2} \label{Lag2}.
\end{gather}

For an integer $a$ not divisible by $2n$, the following identity holds:
\begin{equation}
\label{Id1}
\sum_{m=0}^{n-1} \cos\frac{(2m+1)a\pi}{2n} =0.
\end{equation}

For an integer $a$ not divisible by $2n+1$,
\begin{equation}
\label{Id2}
\sum_{m=0}^{n-1} \cos\frac{(2m+1)a\pi}{2n+1} =\frac{(-1)^{a+1}}{2}
\end{equation}
holds, while for an integer $a$ not divisible by $2n+2$, we have
\begin{equation}
\label{Id3}
\sum_{m=1}^{n} \cos\frac{ma\pi}{n+1} =-\frac{(-1)^{a}+1}{2}
=
\begin{cases}
0 & \text{for odd } a \\
-1 & \text{for even } a.
\end{cases}
\end{equation}

Furthermore, for an integer $a$ not divisible by $2n+1$, we have the following:
\begin{equation}
\label{Id4}
\sum_{m=0}^{n-1} \cos\frac{2(m+1)a\pi}{2n+1} =-\frac{1}{2}.
\end{equation}

\subsection{DCT type 1 case}
\label{sec-sq-dct1}
Elements of the squared transformation matrix are
\begin{align*}
t_{kl} & =\sum_{m=0}^{n-1} \cos\frac{km\pi}{n-1} \cos\frac{ml\pi}{n-1} \\
 & = \frac{1}{2} \sum_{m=0}^{n-1} \cos\frac{m(k+l)\pi}{n-1} + \frac{1}{2} \sum_{m=0}^{n-1} \cos\frac{m(k-l)\pi}{n-1}.
\end{align*}
Using (\ref{Lag1}), for $k\ne l$, we further get
\begin{equation*}
t_{kl} = \frac{1+(-1)^{k+l}}{2}=
\begin{cases}
0 & \text{for odd } k+l \\
1 & \text{for even } k+l.
\end{cases}
\end{equation*}
For $k=l$, we obtain
\begin{equation*}
t_{kk} = 
\begin{cases}
\frac{n+1}{2} & \text{for } k=1,2,\ldots,n-2 \\
n & \text{for } k=0 \text{ or } k=n-1.
\end{cases}
\end{equation*}

\subsection{DCT type 4 case}
In this case, the elements of the squared transformation matrix are
\label{sec-sq-dct4}
\begin{align*}
t_{kl}  ={} & \sum_{m=0}^{n-1} \cos\frac{(2k+1)(2m+1)\pi}{4n} \cos\frac{(2m+1)(2l+1)\pi}{4n} \\
  ={} & \frac{1}{2} \sum_{m=0}^{n-1} \cos\frac{(2m+1)(k+l+1)\pi}{2n} \\
  & + \frac{1}{2} \sum_{m=0}^{n-1} \cos\frac{(2m+1)(k-l)\pi}{2n}.
\end{align*}
For $k\ne l$, using (\ref{Id1}), we get $t_{kl}=0$.
For $k=l$, using (\ref{Id1}) for the first sum only we have
\begin{equation*}
t_{kk}= \frac{1}{2} \sum_{m=0}^{n-1} \cos\frac{(2m+1)(2k+1)\pi}{2n} + \frac{n}{2} = \frac{n}{2}.
\end{equation*}

\subsection{DCT type 5 case}

For this type of the transform we have
\label{sec-sq-dct5}
\begin{align*}
t_{kl}  ={} & \sum_{m=0}^{n-1} \cos\frac{2km\pi}{2n-1} \cos\frac{2ml\pi}{2n-1} \\
  ={} & \frac{1}{2} \sum_{m=0}^{n-1} \cos\frac{2m(k+l)\pi}{2n-1} 
  + \frac{1}{2} \sum_{m=0}^{n-1} \cos\frac{2m(k-l)\pi}{2n-1}.
\end{align*}
For $k\ne l$ using (\ref{Lag2}) we obtain $t_{kl}=1/2$, while for $k=l$ we get
$$t_{kk}=
\begin{cases}
\frac{2n+1}{4} & \text{for } k\ne 0 \\
n & \text{for } k= 0.
\end{cases}
$$

\subsection{DCT type 8 case}
The squared transformation matrix elements for the DCT$_{(8)}$ are given by
\label{sec-sq-dct8}
\begin{align*}
t_{kl}  ={} & \sum_{m=0}^{n-1} \cos\frac{(2k+1)(2m+1)\pi}{4n+2} \cos\frac{(2m+1)(2l+1)\pi}{4n+2} \\
  ={} & \frac{1}{2} \sum_{m=0}^{n-1} \cos\frac{(2m+1)(k+l+1)\pi}{2n+1} \\
  & + \frac{1}{2} \sum_{m=0}^{n-1} \cos\frac{(2m+1)(k-l)\pi}{2n+1}.
\end{align*}
For $k\ne l$, using (\ref{Id2}) we get $t_{kl}=0$.
Otherwise, for $k=l$, using (\ref{Id2}) for the first sum, we get
\begin{equation*}
t_{kk}= \frac{1}{4} + \frac{n}{2} = \frac{2n+1}{4}.
\end{equation*}

\subsection{DST type 1 case}
In the case of DST$_{(1)}$, the elements of the transformation matrix are given by
\label{sec-sq-dst1}
\begin{align*}
t_{kl}  ={} & \sum_{m=0}^{n-1} \sin\frac{(k+1)(m+1)\pi}{n+1} \sin\frac{(m+1)(l+1)\pi}{n+1} \\
  ={} & -\frac{1}{2} \sum_{m=0}^{n-1} \cos\frac{(m+1)(k+l+2)\pi}{n+1} \\
  & + \frac{1}{2} \sum_{m=0}^{n-1} \cos\frac{(m+1)(k-l)\pi}{n+1}.
\end{align*}
For $k\ne l$, using (\ref{Id3}), we have $t_{kl}=0$, while
for $k=l$, $t_{kk}=(n+1)/2$ holds.

\subsection{DST type 4 case}
The elements of the squared  DST$_{(4)}$ transformation matrix are
\label{sec-sq-dst4}
\begin{align*}
t_{kl}  ={} & \sum_{m=0}^{n-1} \sin\frac{(2k+1)(2m+1)\pi}{4n} \sin\frac{(2m+1)(2l+1)\pi}{4n} \\
  ={} & -\frac{1}{2} \sum_{m=0}^{n-1} \cos\frac{(2m+1)(k+l+1)\pi}{2n} \\
  & + \frac{1}{2} \sum_{m=0}^{n-1} \cos\frac{(2m+1)(k-l)\pi}{2n}.
\end{align*}
For $k\ne l$, using (\ref{Id1}) we have $t_{kl}=0$, and for $k=l$ we get $t_{kk}=n/2$.

\subsection{DST type 5 case}
Next, we consider the elements of the squared DST$_{(5)}$ transformation matrix. These elements are given by
\label{sec-sq-dst5}
\begin{align*}
t_{kl}  ={} & \sum_{m=0}^{n-1} \sin\frac{2(k+1)(m+1)\pi}{2n+1} \sin\frac{2(m+1)(l+1)\pi}{2n+1} \\
  ={} & -\frac{1}{2} \sum_{m=0}^{n-1} \cos\frac{2(m+1)(k+l+2)\pi}{2n+1} \\
  & + \frac{1}{2} \sum_{m=0}^{n-1} \cos\frac{2(m+1)(k-l)\pi}{2n+1}.
\end{align*}
For $k\ne l$, using (\ref{Id4}) we have $t_{kl}=0$, and for $k=l$ we get $t_{kk}=(2n+1)/4$.

\subsection{DST type 8 case}
For the DST$_{(8)}$ the elements of squared  matrix are
\label{sec-sq-dst8}
\begin{align*}
t_{kl}  ={} & \sum_{m=0}^{n-1} \sin\frac{(2k+1)(2m+1)\pi}{4n-2} \sin\frac{(2m+1)(2l+1)\pi}{4n-2} \\
  ={} & -\frac{1}{2} \sum_{m=0}^{n-1} \cos\frac{(2m+1)(k+l+1)\pi}{2n-1} \\
  & + \frac{1}{2} \sum_{m=0}^{n-1} \cos\frac{(2m+1)(k-l)\pi}{2n-1}.
\end{align*}
In this case, for $k\ne l$, using (\ref{Id2}) we get $t_{kl}=(-1)^{k+l}/2$, whereas for $k=l$ we have
\begin{equation*}
t_{kk}=
\begin{cases}
\frac{2n+1}{4} & \text{for } k=0,1,\ldots, n-2 \\
n & \text{for } k= n-1.
\end{cases}
\end{equation*}

\section{Trace of transformation matrix}
\label{sec-traces}
In this section, we will determine the trace for each of the previous transformation matrices. To this aim, we  use the following identities
\begin{equation}
\label{trId1}
   \sum_{k=0}^{m-1}\cos\frac{2k^2\pi}{m}
=\frac{\sqrt{m}}{2}\left(1+\cos\frac{m\pi}{2}+\sin\frac{m\pi}{2}\right)
\end{equation}
\begin{equation}
\label{trId2}
\sum_{k=0}^{m-1}\sin\frac{2k^2\pi}{m}
=\frac{\sqrt{m}}{2}\left(1+\cos\frac{m\pi}{2}-\sin\frac{m\pi}{2}\right).
\end{equation}

For the DCT$_{(1)}$, DCT$_{(4)}$, DST$_{(1)}$, and DST$_{(4)}$, for an even number, $n$, the diagonal elements, $d_k$, of transformation matrix are anti-symmetric, that is, $d_k=-d_{n-1-k}$. The trace of these matrices is obviously zero (for an even $n$), 
\begin{equation}
\label{tr-even}
\trace{\mathbf{C}_{(1)}}=\trace{\mathbf{C}_{(4)}}=\trace{\mathbf{S}_{(1)}}= \trace{\mathbf{S}_{(4)}}=0, \  \text{for even } n.    
\end{equation}

Next, we derive the value of the transformation matrix trace in other cases. 
 
\subsection{DCT type 1, odd $n$ case}
\label{sec-tr-dct1odd}
From (\ref{trId1}), using $m=2(n-1)$,  we get
\begin{align*}
\sqrt{2(n-1)} & =   \sum_{k=0}^{2n-3}\cos\frac{k^2\pi}{n-1} \\
 & =  \sum_{k=0}^{n-2}\cos\frac{k^2\pi}{n-1} + \sum_{k=0}^{n-2}\cos\frac{(k+n-1)^2\pi}{n-1} \\
 & =  2\sum_{k=0}^{n-2}\cos\frac{k^2\pi}{n-1} = 2\sum_{k=0}^{n-1}\cos\frac{k^2\pi}{n-1} -2.
\end{align*}
Previous result is the basis for the explicit expression for the trace of the DCT type 1, $\mathbf{C}_{(1)}$, given as follows
\begin{equation}
\label{tr-dct1}
\trace{\mathbf{C}_{(1)}}   = \sum_{k=0}^{n-1} \cos \frac{k^2\pi}{n-1} =  
\begin{cases}
0 & \text{for even } n \\
\frac{2+\sqrt{2n-2}}{2} & \text{for odd } n.
\end{cases}
\end{equation}

\subsection{DCT type 4, odd $n$ case}
\label{sec-tr-dct4odd}
Substituting $m=8n$ in (\ref{trId1}) we get
\begin{align*}
2\sqrt{2n} & =   \sum_{k=0}^{8n-1}\cos\frac{k^2\pi}{4n} \\
 & =  \sum_{k=0}^{4n-1}\cos\frac{k^2\pi}{4n} + \sum_{k=0}^{4n-1}\cos\frac{(k+4n)^2\pi}{4n} \\
 &= 2 \sum_{k=0}^{4n-1}\cos\frac{k^2\pi}{4n} \\
 &= 2 \sum_{k=0}^{2n-1}\cos\frac{k^2\pi}{n} +  2\sum_{k=0}^{2n-1}\cos\frac{(2k+1)^2\pi}{4n}.
\end{align*}
The first sum follows from (\ref{trId1}) when we put $m=2n$, and since $n$ is odd it is equal to zero. Let us decompose the second sum as
\begin{align*}
2\sqrt{2n} = &{} 2\sum_{k=0}^{n-1}\cos\frac{(2k+1)^2\pi}{4n} 
 + 2\sum_{k=0}^{n-1}\cos\frac{(2n+2k+1)^2\pi}{4n} \\
 = {}&{} 4 \sum_{k=0}^{n-1}\cos\frac{(2k+1)^2\pi}{4n}.
\end{align*}
Now we get the trace for this transformation matrix as
\begin{equation}
\label{tr-dct4}
\trace{\mathbf{C}_{(4)}} = \sum_{k=0}^{n-1} \cos \frac{(2k+1)^2\pi}{4n}=
\begin{cases}
0 & \text{for even } n \\
\sqrt{\frac{n}{2}} & \text{for odd } n.
\end{cases}
\end{equation}

\subsection{DCT type 5 case}
\label{sec-tr-dct5} 
For the DCT$_{(5)}$ we can use (\ref{trId1}), with $m=2n-1$, to obtain
\begin{gather*}
\frac{\sqrt{2n-1}}{2}(1+(-1)^n)  = \sum_{k=0}^{2n-2}\cos\frac{2k^2\pi}{2n-1} \\
 = \sum_{k=0}^{n-1}\cos\frac{2k^2\pi}{2n-1} + \sum_{k=0}^{n-2}\cos\frac{2(2n-2-k)^2\pi}{2n-1} \\
 = \sum_{k=0}^{n-1}\cos\frac{2k^2\pi}{2n-1} + \sum_{k=0}^{n-2}\cos\frac{2(k+1)^2\pi}{2n-1} \\
 = 2\sum_{k=0}^{n-1}\cos\frac{2k^2\pi}{2n-1} -1.
\end{gather*}
Finally, the trace of  $\mathbf{C}_{(5)}$ is given as
\begin{equation}
\label{tr-dct5}
\trace{\mathbf{C}_{(5)}} = \sum_{k=0}^{n-1} \cos \frac{2k^2\pi}{2n-1}=
\begin{cases}
\frac{1}{2} & \text{for even } n \\
\frac{1+\sqrt{2n-1}}{2} & \text{for odd } n.
\end{cases}
\end{equation}

\subsection{DCT type 8 case}
\label{sec-tr-dct8} 
For the DCT$_{(8)}$ we can set $m=8n+4$ into (\ref{trId1}), further leading to
\begin{align*}
\sqrt{8n+4} & = \sum_{k=0}^{8n+3}\cos\frac{k^2\pi}{4n+2} \\
& = \sum_{k=0}^{4n+1}\cos\frac{k^2\pi}{4n+2} + \sum_{k=0}^{4n+1}\cos\frac{(k+4n+2)^2\pi}{4n+2} \\
& = 2 \sum_{k=0}^{4n+1}\cos\frac{k^2\pi}{4n+2} \\
& = 2 \sum_{k=0}^{2n}\cos\frac{2k^2\pi}{2n+1} + 2 \sum_{k=0}^{2n}\cos\frac{(2k+1)^2\pi}{4n+2} \\
& = 2 \sum_{k=0}^{2n}\cos\frac{2k^2\pi}{2n+1} +
2\sum_{k=0}^{n-1}\cos\frac{(2k+1)^2\pi}{4n+2} \\
& \quad + 2\sum_{k=0}^{n}\cos\frac{(2(k+n)+1)^2\pi}{4n+2}.
\end{align*}
The first sum is equal to
$$
\sum_{k=0}^{2n}\cos\frac{2k^2\pi}{2n+1} = \frac{\sqrt{2n+1}}{2}(1+(-1)^n),
$$
according to (\ref{trId1}) with $m=2n+1$.

We have that the third sum can be written as
\begin{gather*}
\sum_{k=0}^{n}\cos\frac{(2(k+n)+1)^2\pi}{4n+2} =
\sum_{k=0}^{n}\cos\left(\frac{2k^2\pi}{2n+1}+\frac{2n+1}{2}\pi \right) \\
=(-1)^{n+1}\sum_{k=0}^{n}\sin\frac{2k^2\pi}{2n+1}.
\end{gather*}
From (\ref{trId2}), using $m=2n+1$, we get
\begin{gather*}
\frac{\sqrt{2n+1}}{2}(1-(-1)^n) = \sum_{k=0}^{2n}\sin\frac{2k^2\pi}{2n+1} \\
= \sum_{k=0}^{n}\sin\frac{2k^2\pi}{2n+1} + \sum_{k=0}^{n-1}\sin\frac{2(2n-k)^2\pi}{2n+1} \\
= \sum_{k=0}^{n}\sin\frac{2k^2\pi}{2n+1} + \sum_{k=0}^{n-1}\sin\frac{2(k+1)^2\pi}{2n+1} = 2 \sum_{k=0}^{n}\sin\frac{2k^2\pi}{2n+1}. 
\end{gather*}

Now we have
\begin{gather*}
\sqrt{8n+4} = \sqrt{2n+1}(1+(-1)^n) \\
+ 2 \sum_{k=0}^{n-1}\cos\frac{(2k+1)^2\pi}{4n+2} +   \frac{\sqrt{2n+1}}{2}(1-(-1)^{n}) 
\end{gather*}
or
\begin{gather*}
\sum_{k=0}^{n-1}\cos\frac{(2k+1)^2\pi}{4n+2} =
\frac{\sqrt{2n+1}}{4}(1-(-1)^n) 
\end{gather*}
resulting in the trace of the DCT$_{(8)}$ of the form
\begin{equation}
\label{tr-dct8}
\trace{\mathbf{C}_{(8)}} = \sum_{k=0}^{n-1} \cos \frac{(2k+1)^2\pi}{4n+2} =
\begin{cases}
0 & \!\!\text{for even } n \\
\frac{\sqrt{2n+1}}{2} & \!\!\text{for odd } n.
\end{cases}
\end{equation}
 
\subsection{DST type 1, odd $n$ case}
\label{sec-tr-dst1odd} 
Using $m=2(n+1)$ in \ref{trId2} we get
\begin{align*}
\sqrt{2(n+1)} & = \sum_{k=0}^{2n+1}\sin\frac{k^2\pi}{n+1}
=  \sum_{k=0}^{2n}\sin\frac{(k+1)^2\pi}{n+1}\\
& = \sum_{k=0}^{n-1}\sin\frac{(k+1)^2\pi}{n+1} + \sum_{k=0}^{n}\sin\frac{(n+1+k)^2\pi}{n+1} \\
& = \sum_{k=0}^{n-1}\sin\frac{(k+1)^2\pi}{n+1} + \sum_{k=0}^{n}\sin\frac{k^2\pi}{n+1} \\
& = \sum_{k=0}^{n-1}\sin\frac{(k+1)^2\pi}{n+1} + \sum_{k=0}^{n-1}\sin\frac{(k+1)^2\pi}{n+1} \\
& = 2 \sum_{k=0}^{n-1}\sin\frac{(k+1)^2\pi}{n+1}.
\end{align*}
The trace of this transformation matrix is then
\begin{equation}
\label{tr-dst1}
\trace{\mathbf{S}_{(1)}} = \sum_{k=0}^{n-1} \sin \frac{(k+1)^2\pi}{n+1} =
\begin{cases}
0 & \text{for even } n \\
\sqrt{\frac{n+1}{2}} & \text{for odd } n.
\end{cases}
\end{equation}

\subsection{DST type 4, odd $n$ case}
\label{sec-tr-dst4odd}
Let start from (\ref{trId2}) with $m=8n$, to get
\begin{align*}
2\sqrt{2n} & = \sum_{k=0}^{8n-1}\sin\frac{k^2\pi}{4n} \\
& =  \sum_{k=0}^{4n-1}\sin\frac{k^2\pi}{4n} + \sum_{k=0}^{4n-1}\sin\frac{(4n+k)^2\pi}{4n} \\
& = 2 \sum_{k=0}^{4n-1}\sin\frac{k^2\pi}{4n} \\
& = 2 \sum_{k=0}^{2n-1}\sin\frac{2k^2\pi}{2n} +
2 \sum_{k=0}^{2n-1}\sin\frac{(2k+1)^2\pi}{4n}.
\end{align*}
Using (\ref{trId2}) with $m=2n$ and having in mind that $n$ is odd, we have
\begin{align*}
2\sqrt{2n} & = 2 \sum_{k=0}^{2n-1}\sin\frac{(2k+1)^2\pi}{4n} \\
& = 2 \sum_{k=0}^{n-1}\sin\frac{(2k+1)^2\pi}{4n} + 2 \sum_{k=0}^{n-1}\sin\frac{(2n+2k+1)^2\pi}{4n} \\
& = 4 \sum_{k=0}^{n-1}\sin\frac{(2k+1)^2\pi}{4n}.
\end{align*}
Now we get the trace of DST$_{(4)}$ as
\begin{equation}
\label{tr-dst4}
\trace{\mathbf{S}_{(4)}} = \sum_{k=0}^{n-1} \sin \frac{(2k+1)^2\pi}{4n} =
\begin{cases}
0 & \text{for even } n \\
\sqrt{\frac{n}{2}} & \text{for odd } n.
\end{cases}
\end{equation}

\subsection{DST type 5 case}
\label{sec-tr-dst5}
Using equation (\ref{trId2}) with $m=2n+1$ we can write
\begin{gather*}
\frac{\sqrt{2n+1}}{2} (1-(-1)^n)  = \sum_{k=0}^{2n}\sin\frac{2k^2\pi}{2n+1} \\
=\sum_{k=0}^{2n-1}\sin\frac{2(k+1)^2\pi}{2n+1} \\
= \sum_{k=0}^{n-1}\sin\frac{2(k+1)^2\pi}{2n+1} +
\sum_{k=0}^{n-1}\sin\frac{2(2n-k)^2\pi}{2n+1} \\
= 2\sum_{k=0}^{n-1}\sin\frac{2(k+1)^2\pi}{2n+1}.
\end{gather*}
The trace for the DST$_{(5)}$ is now obtained in the following explicit form
\begin{equation}
\label{tr-dst5}
\trace{\mathbf{S}_{(5)}} = \sum_{k=0}^{n-1} \sin \frac{2(k+1)^2\pi}{2n+1}=
\begin{cases}
0 & \!\!\text{for even } n \\
\frac{\sqrt{2n+1}}{2} & \!\!\text{for odd } n.
\end{cases}
\end{equation}

\subsection{DST type 8}
\label{sec-tr-dst8}
Equation (\ref{trId2}), with $m=8n-4$, can be transformed in the following way
\begin{align*}
\sqrt{8n-4} & = \sum_{k=0}^{8n-5}\sin\frac{k^2\pi}{4n-2} \\
& = \sum_{k=0}^{4n-3}\sin\frac{k^2\pi}{4n-2} + \sum_{k=0}^{4n-3}\sin\frac{(4n-2+k)^2\pi}{4n-2} \\
& =  2\sum_{k=0}^{4n-3}\sin\frac{k^2\pi}{4n-2} \\
& = 2 \sum_{k=0}^{2n-2}\sin\frac{2k^2\pi}{2n-1} + 2 \sum_{k=0}^{2n-2}\sin\frac{(2k+1)^2\pi}{4n-2} \\
& = 2 \sum_{k=0}^{2n-2}\sin\frac{2k^2\pi}{2n-1} + 2 \sum_{k=0}^{n-1}\sin\frac{(2k+1)^2\pi}{4n-2} \\
& \quad + 
2 \sum_{k=0}^{n-2}\sin\frac{(2(k+n)+1)^2\pi}{4n-2} \\
& = 2 \sum_{k=0}^{2n-2}\sin\frac{2k^2\pi}{2n-1} + 2 \sum_{k=0}^{n-1}\sin\frac{(2k+1)^2\pi}{4n-2} \\
& \quad - 2 (-1)^n\sum_{k=0}^{n-2}\cos\frac{2(k+1)^2\pi}{2n-1}.
\end{align*}
The first sum can be calculated using (\ref{trId2}) with $m=2n-1$ as
\begin{equation*}
2\sum_{k=0}^{2n-2}\sin\frac{2k^2\pi}{2n-1} = \sqrt{2n-1}(1+(-1)^n).
\end{equation*}
The third sum follows from (\ref{trId1}) with $m=2n-1$ as
\begin{gather*}
\frac{\sqrt{2n-1}}{2}(1-(-1)^n) = \sum_{k=0}^{2n-2}\cos\frac{2k^2\pi}{2n-1} \\
= 1 + \sum_{k=0}^{n-2}\cos\frac{2(k+1)^2\pi}{2n-1} + \sum_{k=0}^{n-2}\cos\frac{2(2n-2-k)^2\pi}{2n-1} \\
= 1 + 2 \sum_{k=0}^{n-2}\cos\frac{2(k+1)^2\pi}{2n-1}.
\end{gather*}
Now, we can write
\begin{gather*}
\sqrt{8n-4} = \sqrt{2n-1}(1+(-1)^n) + 2\sum_{k=0}^{n-1}\sin\frac{(2k+1)^2\pi}{4n+2} \\ -(-1)^n\left(\frac{\sqrt{2n-1}}{2}(1-(-1)^n) -1\right), 
\end{gather*}
resulting in
\begin{gather*}
\sum_{k=0}^{n-1}\sin\frac{(2k+1)^2\pi}{4n+2}= \frac{\sqrt{2n-1}}{4}(1-(-1)^n)-\frac{(-1)^n}{2}, 
\end{gather*}
and finally
\begin{equation}
\label{tr-dst8}
\trace{\mathbf{S}_{(8)}} = \sum_{k=0}^{n-1} \sin \frac{(2k+1)^2\pi}{4n-2}=
\begin{cases}
-\frac{1}{2} & \text{for even } n \\
\frac{1+\sqrt{2n-1}}{2} & \text{for odd } n.
\end{cases}
\end{equation}


\section{Proofs for DTT eigenvalues and their multiplicities}
\label{sec-Proofs}

Within this section we will use notation $\mathbf{e}_k$ for the standard basis vectors, that is, $\mathbf{e}_k$, $k=1,2,\ldots,n$ is the $k$-th column of $n\times n$ identity matrix.

\subsection{DTT with two eigenvalues}
Here we will analyze the DCT of type 4 and 8 and the DST of type 1, 4 and 5. In all considered cases, according to results presented in Sections \ref{sec-sq-dct4}, \ref{sec-sq-dct8}, \ref{sec-sq-dst1}, \ref{sec-sq-dst4} and \ref{sec-sq-dst5}, square of the transformation matrix is proportional to the identity matrix $\mathbf{I}$, 
\begin{align*}
\mathbf{C}_{(4)}^2 = \mathbf{S}_{(4)}^2 & = \frac{n}{2} \mathbf{I} \\
\mathbf{C}_{(8)}^2 = \mathbf{S}_{(5)}^2& = \frac{2n+1}{4} \mathbf{I} \\
\mathbf{S}_{(1)}^2 & = \frac{n+1}{2} \mathbf{I},
\end{align*}
resulting in eigenvalues of the corresponding matrices as in Table \ref{tab-ev}. Multiplicity of the positive and negative eigenvalue can be determined by calculating the trace of the transformation matrix. Denote by ${{p}}$ the multiplicity of positive eigenvalue and by ${{m}}$ the multiplicity of negative eigenvalue. We know that ${{p}} + {{m}} =n$ and that the trace of the transformation matrix is $({{p}} - {{m}})\lambda^+$, where  $\lambda^+$ is the positive eigenvalue.

Using (\ref{tr-dct4}), (\ref{tr-dct8}), (\ref{tr-dst1}), (\ref{tr-dst4}), and (\ref{tr-dst5}) we can conclude that
for even $n$, the traces of the transformation matrices are equal to zero, meaning that ${{p}}={{m}}=n/2$. For odd $n$ in each case we obtain that the trace of the transformation matrix is $\lambda^+$, meaning that ${{p}}={{m}} +1$, that is, ${{p}}=(n+1)/2$ and ${{m}} = (n-1)/2$.

\subsection{DCT type 5 case}
Let us consider the following decomposition of $\mathbb{R}^n$
\begin{align*}
  \mathbb{R}^n&=\mathcal{V}_1 \oplus \mathcal{V}_2, 
\end{align*}
where vectors in $\mathcal{V}_1$ are of the form
$$
\mathbf{v}=[0,v_1,v_2,\ldots,v_{n-1}]^T
$$
with
$$
\sum_{k=1}^{n-1} v_k = 0
$$
and $\mathcal{V}_2$ is a two-dimensional space generated by
\begin{align*}
\mathbf{q}_1 & =[1,0,0,\ldots,0]^T = \mathbf{e}_1 \\
\mathbf{q}_2 & =[0,1,1,\ldots,1]^T = \sum_{k=2}^{n} \mathbf{e}_{k}.
\end{align*}

The square of  the DCT type 5 transform matrix, $\mathbf{C}_{(5)}$, according to the results presented in Section \ref{sec-sq-dct5}, can be written as
\begin{equation}
\label{dct5-a}
 \mathbf{C}_{(5)}^2 =\diag_n\left(\frac{2n-1}{2}, \frac{2n-1}{4},\ldots,\frac{2n-1}{4}\right)+\frac{1}{2}\mathbf{1}_n,
\end{equation}
where $\diag_n(\cdot)$ is an $n\times n$ diagonal matrix and $\mathbf{1}_n$ is an $n\times n$ matrix with all ones. It is obvious that $\mathcal{V}_1$ is within null space of $\mathbf{1}_n$. Now we have
\begin{equation}
\mathbf{C}_{(5)}^2 \mathbf{v} = \frac{2n-1}{4} \mathbf{v} \qquad (\forall \mathbf{v} \in \mathcal{V}_1),
\label{dct5v1}
\end{equation}
meaning that $\frac{2n-1}{4}$ is an eigenvalue of $\mathbf{C}_{(5)}^2$. {Note that both  of the following matrices 
\begin{equation}
\label{dct5-a1}
\mathbf{C}_{(5)}\pm\sqrt{\frac{2n-1}{4}}\mathbf{I}
\end{equation}
are singular.

If $\mathbf{C}_{(5)}-\sqrt{\frac{2n-1}{4}}\mathbf{I}$ is non-singular, based on (\ref{dct5v1}), we have
$$\mathbf{C}_{(5)}\mathbf{v}=\sqrt{\frac{2n-1}{4}}\mathbf{v} \qquad (\forall \mathbf{v} \in \mathcal{V}_1),$$
which is impossible. It implies that $\lambda = \pm \sqrt{\frac{2n-1}{4}}$ are  eigenvalues of $\mathbf{C}_{(5)}$. Moreover the sum of the multiplicities is at least $n-2$.} 
{ Since  $\mathcal{V}_1$ is invariant under the symmetric transform  $\mathbf{C}_{(5)}$,  the other summand  $\mathcal{V}_2$ remains invariant as well. Thus, focusing on $\mathcal{V}_2$, we will find the other eigenvalues.}  We have that
\begin{align}
\mathbf{C}_{(5)}\mathbf{q}_1 & =\mathbf{q}_1+\mathbf{q}_2 \label{dct5-aa} \\
\mathbf{C}_{(5)}\mathbf{q}_2 & =(n-1)\mathbf{q}_1-\frac{1}{2}\mathbf{q}_2. \label{dct5-bb}
\end{align}
The eigenvalues (and the corresponding eigenvectors) can be found by solving
$$
\mathbf{C}_{(5)}(\mathbf{q}_1+ x \mathbf{q}_2)=\lambda( \mathbf{q}_1+ x \mathbf{q}_2)
$$
for unknown $\lambda$ (and $x$). Using (\ref{dct5-aa}) and (\ref{dct5-bb}) we get
$$
(x(n-1)+1)\mathbf{q}_1+\mathbf{q}_2 + (1-\frac{x}{2})\mathbf{q}_2=\lambda\mathbf{q}_1+ \lambda x \mathbf{q}_2,
$$
resulting in a system of equations
\begin{align*}
x(n-1)+1 & = \lambda \\
1-\frac{x}{2} & = \lambda x,
\end{align*}
and the eigenvalues
\begin{equation*}
\lambda = \frac{1}{4} \pm \sqrt{n-\frac{7}{16}},
\end{equation*}
each with multiplicity one, as stated in Table \ref{tab-ev}. {Therefore, we can conclude that $\mathbf{C}_{(5)}$ has just four distinct eigenvalues.} 

Multiplicity of eigenvalues can be determined using the trace of the transformation matrix (\ref{tr-dct5}) and the fact that their multiplicities sum to $n-2$. The sum of all eigenvalues is 
$$ \frac{1}{2}+({{p}}-{{m}})\sqrt{\frac{2n-1}{4}}, $$ 
resulting in ${{p}}={{m}}=n/2$ for an even $n$ and ${{p}}=(n-1)/2$, ${{m}}=(n-3)/2$ for an odd $n$.

\subsection{DST type 8 case}
In analogy to the previous case, the subspace $\mathcal{V}_1$ is a set of vectors
$$
\mathbf{v}=[v_0,v_1,v_2,\ldots,v_{n-2},0]^T,
$$
satisfying 
$$
\sum_{k=0}^{n-2} (-1)^{k} v_k = 0
,$$
whereas $\mathcal{V}_2$ is a two-dimensional space generated by
\begin{align*}
\mathbf{q}_1 & =[0,0,\ldots,0,1]^T = \mathbf{e}_n \\
\mathbf{q}_2 & =[1,-1,1,-1,\ldots,(-1)^{n-2},0]^T = -\sum_{k=1}^{n-1} (-1)^{k} \mathbf{e}_{k}.
\end{align*}

The square of $\mathbf{S}_{(8)}$ matrix, according to Section \ref{sec-sq-dst8}, can be written as
\begin{equation}
\label{dst8-S2}
\mathbf{S}_{(8)}^2 =\diag_n\left(\frac{2n-1}{4}, \frac{2n-1}{4},\cdots,\frac{2n-1}{2}\right)+\frac{1}{2}\mathbf{Q}.
\end{equation}
The elements of matrix $\mathbf{Q}$ are defined by $q_{kl}=(-1)^{kl}$. Again, the vector space $\mathcal{V}_1$ is within the null space of $\mathbf{Q}$ resulting in
\begin{equation*}
\mathbf{S}_{(8)}^2 \mathbf{v} = \frac{2n-1}{4} \mathbf{v} \qquad (\forall \mathbf{v} \in \mathcal{V}_1).
\end{equation*}
Therefore, $\frac{2n-1}{4}$ is an eigenvalue of  $\mathbf{S}_{(8)}^2$. One may conclude that both  $\pm \sqrt{\frac{2n-1}{4}}$ are eigenvalues of $\mathbf{S}_{(8)}$. Moreover,  the sum of  multiplicities is at least  $n-2$.

The other  eigenvalues can be found by solving
\begin{equation*}
\mathbf{S}_{(8)} (\mathbf{q}_{1}+ x \mathbf{q}_{2}) = \lambda (\mathbf{q}_{1}+ x \mathbf{q}_{2}).
\end{equation*}
Using
\begin{align*}
      \mathbf{S}_{(8)}\mathbf{q}_1 & =(-1)^{n-1}\mathbf{q}_{1}+\mathbf{q}_{2} \\
      \mathbf{S}_{(8)}\mathbf{q}_2 & =(n-1)\mathbf{q}_{1} +\frac{(-1)^{n}}{2}\mathbf{q}_{2},      
\end{align*}
we obtain the system of equations
\begin{align*}
(-1)^{n-1} +(n-1) x & = \lambda \\
1+\frac{(-1)^n}{2} x & = \lambda x
\end{align*}
with solutions
$$
\lambda = -\frac{(-1)^{n}}{4} \pm \sqrt{n-\frac{7}{16}}.
$$
Each eigenvalue has multiplicity one.

Now we can determine the  multiplicities of all eigenvalues. Denoting by ${{p}}$ the  multiplicity of eigenvalue $\sqrt{(2n-1)/{4}}$ and with ${{m}}$ the multiplicity of $-\sqrt{(2n-1)/{4}}$, where ${{p}} + {{m}} = n-2$ we have that the sum of all eigenvalues is
\begin{equation}
({{p}} - {{m}})\sqrt{\frac{2n-1}{4}} -\frac{(-1)^n}{2}.
\end{equation}
This sum is equal to the trace of  $\mathbf{S}_{(8)}$ matrix (\ref{tr-dst8}). For even $n$ we have that the trace is $-\frac{1}{2}$, meaning that ${{p}} = {{m}} = \frac{n-2}{2}$. For an odd $n$ we have that ${{p}} - {{m}} = 1$, resulting in ${{p}} = \frac{n-1}{2}$ and ${{m}} = \frac{n-3}{2}$, as given in Table \ref{tab-ev}.

\subsection{DCT type 1 case}
{ Similar to the  previous types,  to get the eigenvalues of  $\mathbf{C}_{(1)}$,  we   split the problem  to some simpler components.  In this case, we have to consider the odd and the even cases separately. Moreover   $\mathbb{R}^n$ is decomposed into three orthogonal subspaces,} 
\begin{equation}\label{DC1}
\mathbb{R}^n=\mathcal{V}_1\oplus \mathcal{V}_2 \oplus \mathcal{V}_3.
\end{equation}
Let   $\mathcal{V}_1$ be the  $(n-4)$-dimensional vector space containing vectors $\mathbf{v}=[0,v_1,v_2,\cdots,v_{n-2},0]^T$ such that the sum of even indexed values is zero and the sum of odd indexed values is also zero
$$ \sum_{k=1}^{(n-3)/2} v_{2k} = 0\quad \text{and} \quad  \sum_{k=1}^{(n-1)/2} v_{2k-1} = 0.$$

Note that, according to Section \ref{sec-sq-dct1}, we have
\begin{equation*}
\mathbf{C}_{(1)}^2
=\diag(n-1,\frac{n-1}{2},\frac{n-1}{2},\cdots,\frac{n-1}{2},n-1)
+\mathbf{P},
\end{equation*}
where $\mathbf{P}$ is an $n\times n$ matrix  with elements 
$$p_{kl}=\frac{1+(-1)^{k+l}}{2}=
\begin{cases}
0 & \text{ for odd } k+l\\
1  & \text{ for even } k+l.
\end{cases}
$$
One may directly check that  $\mathcal{V}_1$ is contained in the null space of $\mathbf{P}$. Applying this point, we get that
 \begin{equation}
 \label{C_1}
\mathbf{C}_{(1)}^2\mathbf{v}=\frac{n-1}{2}\mathbf{v} \qquad  (\forall \mathbf{v}\in \mathcal{V}_1).
\end{equation}
Thus, ${(n-1)}/{2}$ is an eigenvalue of $\mathbf{C}_{(1)}^2$.  Similar to the two previous type transforms,  we conclude that  $\pm \sqrt{(n-1)/2}$ are eigenvalues of $\mathbf{C}_{(1)}$ corresponding to $\lambda_3$ and $\lambda_4$ in Table \ref{tab-DCT1} for even and odd $n$.

Suppose that $n$ is odd. Let us define the
vector space $\mathcal{V}_2$ as a two-dimensional space generated by
\begin{align}
\mathbf{q}_{21} & =[1,0,0,\ldots,0,-1]^T = \mathbf{e}_1 - \mathbf{e}_n \\
\mathbf{q}_{22} & =[0,1,0,1,0,\ldots,0,1,0]^T = \sum_{k=1}^{
(n-1)/2
} \mathbf{e}_{2k}, 
\end{align}
and $\mathcal{V}_3$ as a two-dimensional space generated by
\begin{align}
\mathbf{q}_{31} & =[1,0,0,\ldots,0,1]^T = \mathbf{e}_1 + \mathbf{e}_n \\
\mathbf{q}_{32} & =[0,0,1,0,1,\ldots,1,0,0]^T = \sum_{k=1}^{
(n-3)/2
} \mathbf{e}_{2k+1}.
\end{align}

Consider subspace $\mathcal{V}_2$. We have that \Red{Not obvious}
\begin{align}
\mathbf{C}_{(1)} \mathbf{q}_{21} & = 2 \mathbf{q}_{22} \label{dct1-a}\\
\mathbf{C}_{(1)} \mathbf{q}_{22} & = \frac{n-1}{2} \mathbf{q}_{21} \label{dct1-b}.
\end{align}
Let us now find eigenvalues and eigenvectors within this subspace. We should find $x$ such that
\begin{equation*}
\mathbf{C}_{(1)} (\mathbf{q}_{21} + x \mathbf{q}_{22}) = \lambda (\mathbf{q}_{21} + x \mathbf{q}_{22}) 
\end{equation*}
for some $\lambda$. Then $\mathbf{q}_{21} + x \mathbf{q}_{22}$ is an eigenvector in $\mathcal{V}_2$ and $\lambda$ is the corresponding eigenvalue. Using (\ref{dct1-a}) and (\ref{dct1-b}) we get
\begin{equation*}
2\mathbf{q}_{22} + x \frac{n-1}{2}\mathbf{q}_{21} = \lambda \mathbf{q}_{21} + \lambda x \mathbf{q}_{22},
\end{equation*}
resulting in a system of equations with unknown $x$ and $\lambda$
\begin{align*}
2 & = \lambda x \\
x\frac{n-1}{2} & =\lambda.
\end{align*}
Solutions to this system are the eigenvalues $\lambda=\pm\sqrt{n-1}$ corresponding to $\lambda_1$ and $\lambda_5$ in Table \ref{tab-DCT1} for an odd $n$. Each eigenvalue has multiplicity one.

Consider now $\mathcal{V}_3$ space. Similar to the previous case we get \Red{Not obvious}
\begin{align*}
\mathbf{C}_{(1)} \mathbf{q}_{31} & = 2 \mathbf{q}_{31} + 2 \mathbf{q}_{32}\\
\mathbf{C}_{(1)} \mathbf{q}_{32} & = \frac{n-3}{2} \mathbf{q}_{31} -  \mathbf{q}_{31}.
\end{align*}
Next, we search for eigenvectors and eigenvalues from
\begin{equation*}
\mathbf{C}_{(1)} (\mathbf{q}_{31} + x \mathbf{q}_{32}) = \lambda (\mathbf{q}_{31} + x \mathbf{q}_{32}), 
\end{equation*}
leading to the system of equations
\begin{align*}
2 + \frac{n-3}{2} & = \lambda \\
2-x & =\lambda x,
\end{align*}
with the solutions
$$ \lambda=\frac{1}{2} \pm \sqrt{n-\frac{3}{4}}, $$
corresponding to $\lambda_2$ and $\lambda_6$ in Table \ref{tab-DCT1}. Both of the obtained eigenvalues are with multiplicity one.

Note that in the considered cases we have an analytical form for the corresponding eigenvectors.

Now we can determine multiplicities of all eigenvalues. Denoting by ${{p}}$ the multiplicity of eigenvalue $\sqrt{(n-1)/{2}}$ and by ${{m}}$ the multiplicity of $-\sqrt{(n-1)/{2}}$, where ${{p}} + {{m}} = n-4$, we have that the sum of all eigenvalues is
\begin{equation}
({{p}} - {{m}})\sqrt{\frac{n-1}{2}} +1.
\end{equation}
This sum is equal to the trace of  $\mathbf{C}_{(1)}$ matrix (\ref{tr-dct1}). We can conclude that ${{p}} - {{m}} = 1$, resulting in ${{p}} = \frac{n-3}{2}$ and ${{m}} = \frac{n-5}{2}$, as given in Table \ref{tab-DCT1} for odd $n$ case.
\medskip

Now we will consider the case of an even $n$. In decomposition (\ref{DC1}) $\mathcal{V}_1$ remains the same, while $\mathcal{V}_2$ is now spanned by vectors
\begin{align}
\mathbf{q}_{21} & =[1,0,0,\ldots,0,\sqrt{2}+1] = \mathbf{e}_1 + (\sqrt{2}+1) \mathbf{e}_n\\
\mathbf{q}_{22} & =[0,\sqrt{2}+1,1,\sqrt{2}+1,1,\ldots,\sqrt{2}+1,0] \nonumber \\
& = \sum_{k=2}^{n-1} \mathbf{e}_{k} + \sqrt{2} \sum_{k=1}^{
(n-1)/2
} \mathbf{e}_{2k}
\end{align}
and $\mathcal{V}_3$ is spanned by
\begin{align}
\mathbf{q}_{31} & =[1,0,0,\ldots,0,-\sqrt{2}-1] = \mathbf{e}_1 - (\sqrt{2}+1) \mathbf{e}_n\\
\mathbf{q}_{32} & =[0,1-\sqrt{2},1,1-\sqrt{2},1,\ldots,1-\sqrt{2},0] \nonumber\\
& = \sum_{k=2}^{n-1} \mathbf{e}_{k} - \sqrt{2} \sum_{k=1}^{
(n-1)/2
} \mathbf{e}_{2k}.
\end{align}
It is easy to check that $\mathcal{V}_2$ and $\mathcal{V}_3$ are invariant with respect to $\mathbf{C}_{(1)}$, that is, \Red{Not obvious}
\begin{align}
\mathbf{C}_{(1)} \mathbf{q}_{21} & =\sqrt2 \mathbf{q}_{21}+
(2-\sqrt{2}) \mathbf{q}_{22} \label{c1a} \\
\mathbf{C}_{(1)} \mathbf{q}_{22} & =(n-2)(1+\frac{\sqrt2}{2}) \mathbf{q}_{21}-\frac{\sqrt2}{2} \mathbf{q}_{22} \label{c1b}\\
\mathbf{C}_{(1)} \mathbf{q}_{31} & =-\sqrt2 \mathbf{q}_{31}+
(2+\sqrt{2}) \mathbf{q}_{32} \label{c1c} \\
\mathbf{C}_{(1)} \mathbf{q}_{32} & =-(n-2)\frac{\sqrt2}{2} \mathbf{q}_{31}+\frac{\sqrt2}{2} \mathbf{q}_{32}. \label{c1d}
\end{align}
The eigenvalues (with corresponding eigenvectors) can be found by solving the system
\begin{align}
\mathbf{C}_{(1)}(\mathbf{q}_{21} + x \mathbf{q}_{22})  & =\lambda (\mathbf{q}_{21} + x \mathbf{q}_{22}) \label{dct1-e1} \\
\mathbf{C}_{(1)}(\mathbf{q}_{31} + x \mathbf{q}_{32})  & =\lambda (\mathbf{q}_{31} + x \mathbf{q}_{32}), \label{dct1-e2}
\end{align}
for unknown $\lambda$ (and $x$).
Form (\ref{dct1-e1}), using (\ref{c1a}) and (\ref{c1b}), we obtain the system of equations
\begin{align*}
\sqrt{2}+x(n-2)(1+\frac{\sqrt{2}}{2})  & = \lambda \\
2-\sqrt{2} - x\frac{\sqrt{2}}{2} & = \lambda x
\end{align*}
with solutions
\begin{equation*}
\lambda = \frac{\sqrt{2}}{4} \pm \sqrt{n-\frac{7}{8}},
\end{equation*}
corresponding to $\lambda_2$ and $\lambda_6$ in Table \ref{tab-DCT1} for even $n$. In a similar way, by solving (\ref{dct1-e2}) we can obtain $\lambda_1$ and $\lambda_5$.

The sum of all eigenvalues is
\begin{equation}
({{p}} - {{m}})\sqrt{\frac{n-1}{2}},
\end{equation}
and the trace of $\mathbf{C}_{(1)}$ is zero, (\ref{tr-dct1}), resulting in ${{p}} = {{m}} = (n-4)/2$, as stated in Table \ref{tab-DCT1} for an even $n$.

\section{Conclusion}
An analytic proof for eigenvalues, and corresponding multiplicities is provided for eight symmetric non-normalized DTT. The trace and the square of transformation matrix is derived in all analyzed cases. Our further research will include derivation of eigenvector basis for the analyzed DTT. The proposed approach, based on the decomposition of the eigenspace into orthogonal subspaces invariant under considered DTT, provides eigenvalues and some eigenvectors, for the case when the eigenvalue multiplicity is 1. Since the DTT
are fundamental mathematical tools for signal processing and related applications, we believe that the presented theory is particularly relevant to this field, since it sheds a new light on the understanding of commonly used transforms.

\vfill

\end{document}